\documentclass[aps,reprint,showpacs,amsmath,amssymb,amsfonts,prl,superscriptaddress]{revtex4-1}

\usepackage{graphicx}
\usepackage{dcolumn}
\usepackage{bm}
\usepackage{epsf}
\usepackage{amsmath}
\usepackage{amssymb}
\usepackage{amsfonts}
\usepackage{color}

\usepackage{esdiff}	
\usepackage[tight,ugly]{units}	
\usepackage{textcomp}
\usepackage[ansinew]{inputenc}
\usepackage[T1]{fontenc}
\usepackage{lmodern} 
\usepackage{amsfonts}

\newcommand{\bla}{bla\\bla\\bla\\bla\\bla}

\usepackage{setspace}
\usepackage{xspace}	

\newcommand{\dt}{\ensuremath{\text{d}t}\xspace}

\newcommand{\Er}{\ensuremath{E_\text{r}}\xspace}
\newcommand{\pr}{\ensuremath{p_\text{r}}\xspace}

\begin{document}

\title{Infinite  density for cold atoms in shallow optical lattices}

\author{Philip C. Holz}
\affiliation{Institut f{\"u}r Experimentalphysik, Universit{\"a}t Innsbruck, A-6020 Innsbruck, Austria}

\author{Andreas Dechant}
\affiliation{Dahlem Center for Complex Quantum Systems, FU Berlin, D-14195 Berlin, Germany}

\author{Eric Lutz}
\affiliation{Institute for Theoretical Physics, University of Erlangen-N\"urnberg, D-91058 Erlangen, Germany}

\begin{abstract}
Infinite densities can describe the long-time properties of systems when ergodicity is broken and the equilibrium Boltzmann-Gibbs distribution fails. We here perform semiclassical Monte Carlo simulations of cold atoms in dissipative optical lattices with realistic parameters. We show that the  momentum infinite density, as well as its scale invariance, should be observable in shallow potentials. We further evaluate the momentum autocorrelation function in the stationary and aging regime.
\end{abstract}

\maketitle

The Boltzmann-Gibbs distribution plays a pivotal role in physics and mathematics. In statistical mechanics, it gives the probability of finding a system in a given energy state in the canonical ensemble, from which all its equilibrium properties  can be determined \cite{rei65}. In nonequilibrium physics and in the theory of stochastic processes, it appears as the stationary density of Fokker-Planck equations with additive noise \cite{ris89}. On the other hand, in mathematics, the Boltzmann-Gibbs measure is a prominent  example of an invariant probability measure that forms the basis of the statistical description of dynamical systems, such as chaotic nonlinear maps \cite{ott93}. It is also an essential concept in ergodic and probability theory \cite{fel50}. 
It is often assumed that invariant measures are finite, that is, the corresponding probability density is integrable (or normalizable). However, there are dynamical systems of interest whose invariant measures are infinite \cite{zwe09}. In these systems, time averages of observables are intrinsically random and do not converge to the (nonrandom) ensemble averages, even in the limit of long times, in contrast to ergodic systems.  Infinite measures are the subject of infinite ergodic theory \cite{aar97} and have  lately found applications in physics, e.g. in the study of weakly chaotic systems \cite{kor09,kor12,kor13} and subdiffusive maps \cite{aki10,aki12}.

Infinite densities were recently shown to be also of importance  in the investigation of Brownian particles diffusing  in an asymptotically  logarithmic potential \cite{Kes10,dec11a}. This scenario describes  a great variety of systems, including
charged polymers in solution \cite{man69}, self-gravitating Brownian
particles \cite{sir02,cha07}, long-range interacting systems \cite{bou05,cha07a}, and the denaturization
of DNA molecules \cite{fog07,bar07}. It moreover  provides an approximate description of cold atoms in shallow dissipative optical lattices, as discussed in detail below. 
These systems are characterized by a power-law Boltzmann-Gibbs distribution \cite{Lut03} and  algebraic relaxation \cite{Mar96}. For small potential depths, the  Boltzmann-Gibbs distribution fails to account for the long-time properties of the particle, contrary to naive expectations, as it leads to diverging expressions. In this regime, ergodicity is broken and the system therefore lies beyond the reach of statistical mechanics \cite{lut04}. An infinite density for the Brownian particle was obtained as the large  but finite time solution of the corresponding Fokker-Planck equation \cite{Kes10,dec11a}. It possesses the intriguing property that it asymptotically diverges at the origin, and is hence not normalizable. However, its  moments are finite for precisely the parameters for which those of the Boltzmann-Gibbs distribution, the infinite time solution of the Fokker-Planck equation, diverge, and vice versa. The two distributions are thus complementary and  are both needed to predict the observable properties of the particle. The infinite density was further used to evaluate the time-averaged  position of the particle and its two-time autocorrelation function for potential depths for which the Boltzmann-Gibbs distribution cannot be employed \cite{dec12a}. The infinite density therefore appears as an indispensable tool to characterize the asymptotic behavior of the system in the nonergodic phase. However, despite its importance, an infinite probability density has never been experimentally observed.

In this paper, we investigate the possibility to observe an infinite density using cold atoms in dissipative optical lattices \cite{gry01}.
To this end, we perform detailed semiclassical Monte Carlo simulations of the dynamics of the atoms under Sisyphus cooling \cite{Pet99}, using realistic parameters. Cold atoms in optical lattices provide an ideal system to analyze phenomena beyond Boltzmann-Gibbs statistical mechanics \cite{lut13}. Created by counterpropagating laser beams, they can be easily tuned. They were used, for example, to observe the transition from normal to anomalous superdiffusion in shallow potentials in a number of experiments \cite{hod95,kat97,sag12,wic12}.  The atomic dynamics is often described with the help of an approximate Fokker-Planck equation, Eq.~\eqref{3} below, obtained by spatial averaging over one lattice period \cite{Cas90}. The approximation is valid for fast atoms and neglects the contributions from those that are localized in the potential wells. By contrast, the Monte Carlo simulation takes into account the microscopic origin of Sisyphus cooling, including the periodicity of  the lattice potential and transitions between atomic Zeeman sublevels.  The occurrence of a  Boltzmann-Gibbs distribution with tunable power-law tails  predicted by the Fokker-Planck equation was experimentally confirmed in Ref.~\cite{Dou06}, showing that the latter provides a good description of the stationary asymptotic features of the system. However, owing  to the many approximations involved in its derivation, it is unclear whether the Fokker-Planck equation also correctly describes the finite time properties, in particular the infinite density. In the following, we extend the semiclassical Monte Carlo algorithm used for atoms in deep lattices to the shallow lattice regime by removing the appearance of nonphysical divergences. We employ the algorithm to simulate both one-time and two-time functions and compare them with the analytical predictions of the Fokker-Planck equation. We use a recently developed maximum likelihood estimation method with lower and upper cutoffs to determine the parameters of the two functions \cite{Bar12}. Our main results are that the scaling properties  remain unaffected by the approximations entering the Fokker-Planck equation and that the infinite density should therefore be  experimentally observable.  

\textit{Semiclassical approach and infinite density.} 
We consider a 1D optical  lattice created by the superposition of two laser fields with orthogonal linear polarizations (lin$\perp$lin configuration) \cite{gry01}. In the semiclassical approach, the internal degrees of freedom of the atoms are treated quantum  mechanically while the external dynamics  $(x,p)$ is described classically. The starting point of our analysis is the system of two coupled differential equations for the phase space densities $W_{\pm}(x,p,t)$ for the Zeeman sublevels of the atomic ground state $\vert g, m_z = \pm 1/2 \rangle$ derived in Ref.~\cite{Pet99} in the limit of weak laser intensities, 
\begin{align}
\bigg[ \partial_t + &\frac{p}{m} \partial_x - \frac{{\rm d} U_{\pm}(x)}{{\rm d} x} \partial_p \bigg] W_{\pm}(x,p,t) \label{1} \\
& = -\left[\gamma_{\pm \mp}(x) W_{\pm}(x,p,t) - \gamma_{\mp \pm}(x) W_{\mp}(x,p,t)\right] \nonumber \\
& \quad\,+ \partial_p^2 \left[D_{\pm \pm}(x) W_{\pm}(x,p,t) + D_{\mp \pm}(x) W_{\mp}(x,p,t) \right] \nonumber.
\end{align}
The above equations describe the motion of a classical particle with mass $m$ in the periodic potentials 
 $U_{\pm}(x)$.  The  rates $\gamma_{\pm \mp}(x)$ and the diffusion coefficients $D_{\pm \pm}(x)$ and $D_{\mp \pm}(x)$ determine, respectively, the probability of internal transitions and the effect of noise, such as spontaneous emission -- their exact expressions can be found in the Supplemental Material. 
An analytical solution of Eq.~\eqref{1} is not known.
However,  an analytically tractable description can be obtained by averaging the spatial coordinate $x$ over one wavelength. Applying the method described in Ref.~\cite{Cas90} to Eq.~\eqref{1}, we obtain the following Fokker-Planck equation for the momentum probability density $W(p,t) =\int dx\, (W_{+}(x,p,t)+W_{-}(x,p,t))$, 
\begin{align}
\partial_t W(p,t) &= \partial_p \left[ \frac{\gamma p}{1 + (p/p_c)^2} W(p,t) \right] \nonumber \\
& \quad + \partial_p \left[ \left( D_1 + \frac{D_2}{1 + (p/p_c)^2} \right) \partial_p W(p,t) \right] \label{3}.
\end{align}
Equation \eqref{3} is valid for  large momenta, $p \gg \pr=\hbar k$, where  $k$ is the wave number of the light field, and under the assumption that the transitions between the sublevels have reached a steady state, that is, the difference of the sublevel occupations, $\varphi(x,p,t) = W_{+}(x,p,t) - W_{-}(x,p,t)$, is independent of time. 
The coefficients appearing in  Eq.~\eqref{3} are explicitly given by,
\begin{equation}
\gamma \!=\! - \frac{3 \hbar k^2 \delta'}{m \Gamma'}, \, p_c \!=\! \frac{m \Gamma'}{9 k}, \, D_1\! =\! \frac{41 \hbar^2 k^2 \Gamma'}{90}, \, D_2 \!= \!\frac{\hbar^2 k^2 \delta'^2}{\Gamma'} \label{4},
\end{equation}
where $\Gamma' = \Gamma s_0$ and $\delta' = \delta s_0$ are the saturation-adjusted linewidth and detuning, with $s_0$  the saturation parameter and  $\Gamma$  the natural linewidth of the atomic transition. Equation \eqref{3} has the same form as that obtained in previous studies \cite{Mar96,hod95,Cas90}. However, the prefactor of the diffusion coefficient $D_1$  differs from the one given in Ref.~\cite{Cas90} ($41/90$ instead of $11/18$) and explains the noted discrepancy of a factor $4/3$. We notice incidentally that it is equal to the coefficient found empirically in Ref.~\cite{Mar96}. In the limit of large momenta, $p\gg p_c$, considered in most investigations \cite{Kes10,dec11a,Kes12,Dec12}, Eq.~\eqref{3} reduces to a Fokker-Planck equation with inversely linear drift, $\gamma p_c^2/p$, and constant diffusion $D_1$,  that corresponds to Brownian motion in an asymptotically logarithmic potential. 

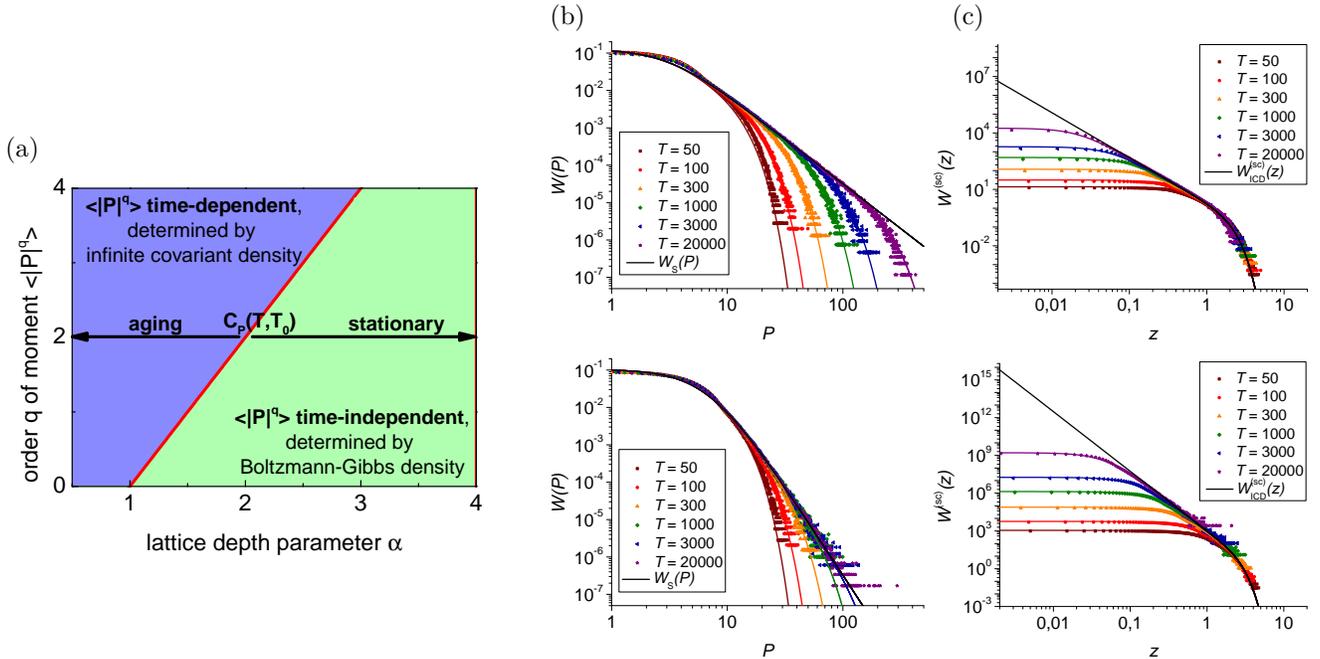
\begin{figure*}[ht]
	\centering	
	\begin{picture}(400,240)
\put(-50,191){(a)}
\put(-65,40){\includegraphics[width=0.44\textwidth]{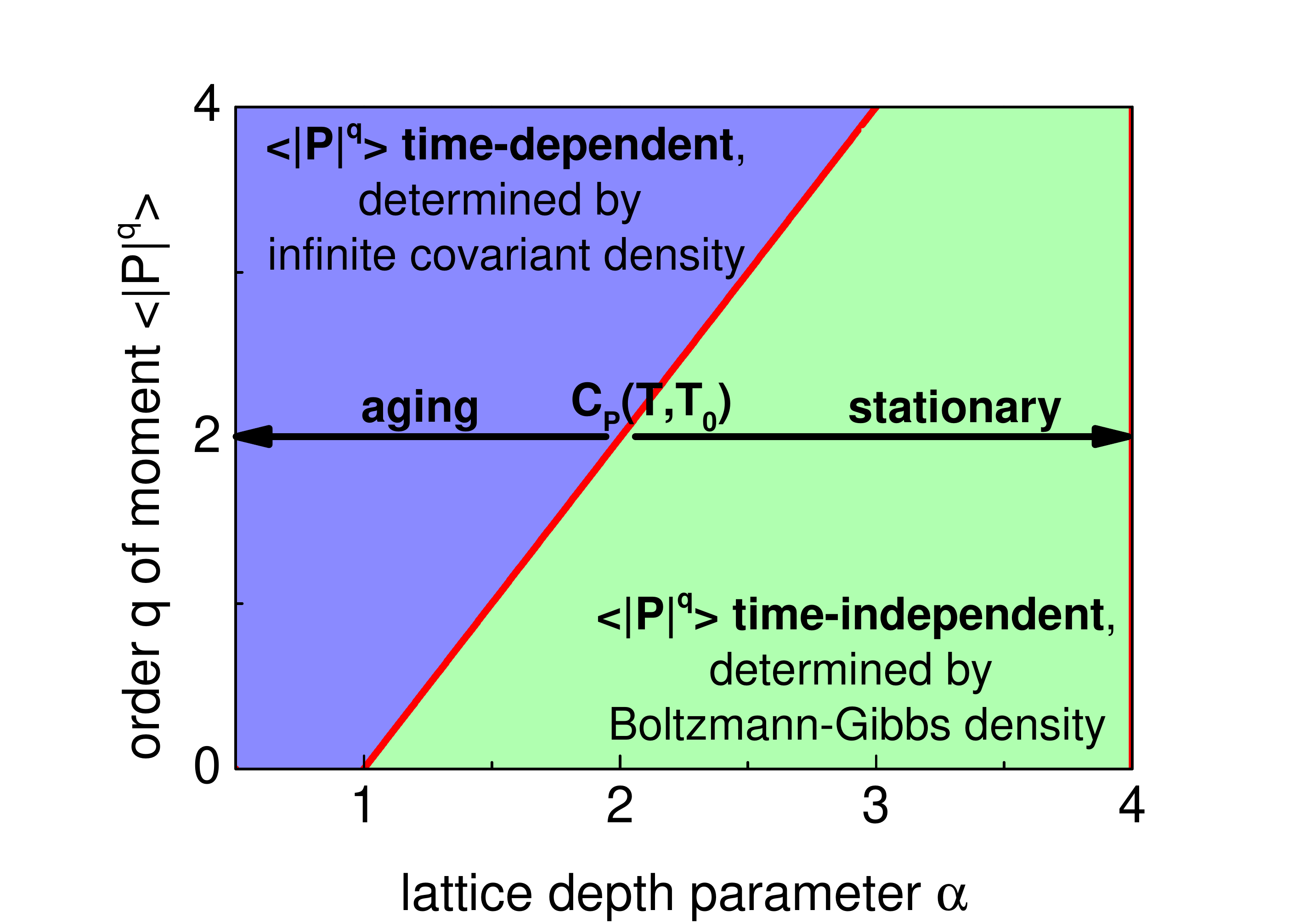}}
\put(156,241){(b)}
\put(155,120){\includegraphics[width=0.28\textwidth]{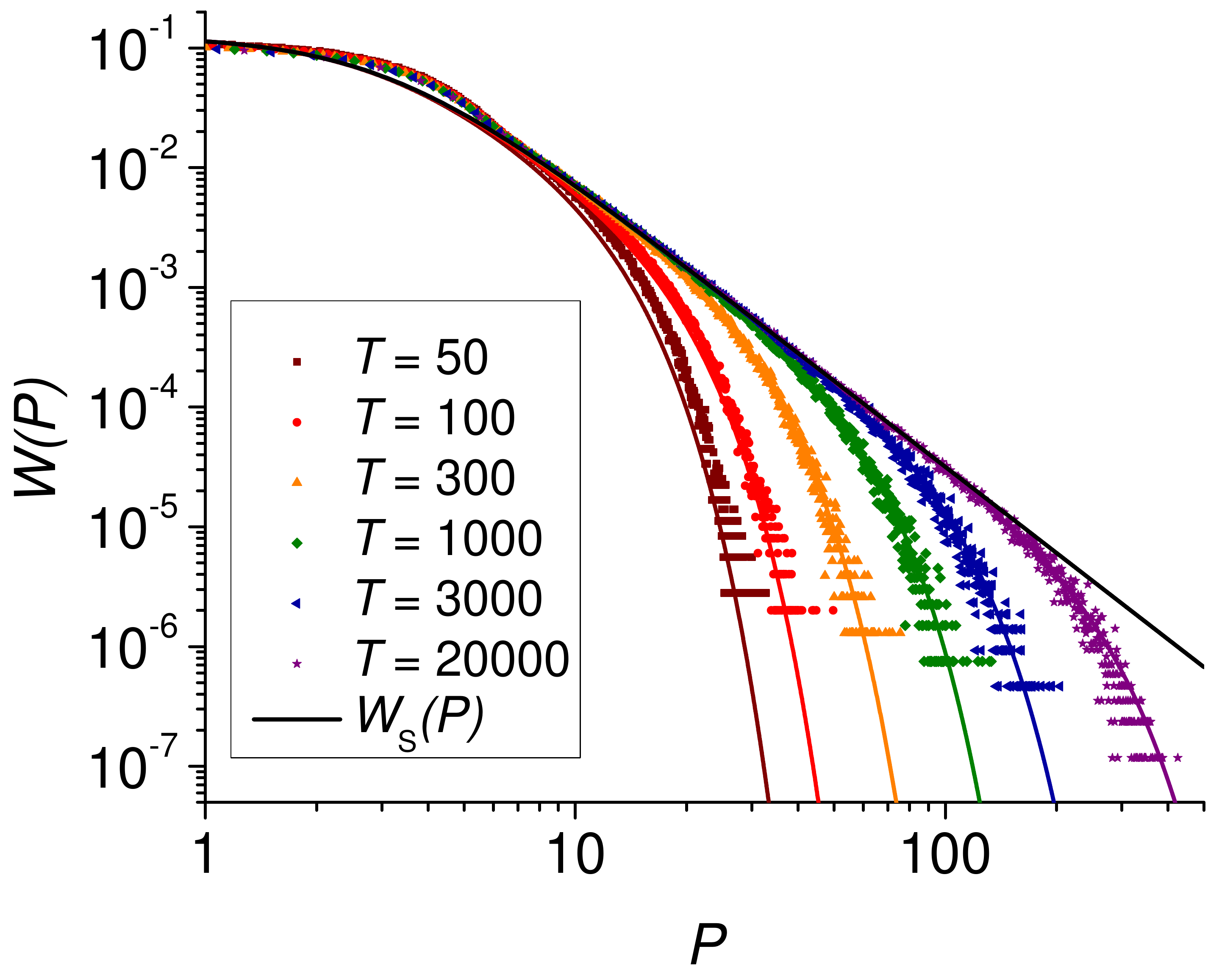}}\put(300,120) {\includegraphics[width=0.28\textwidth]{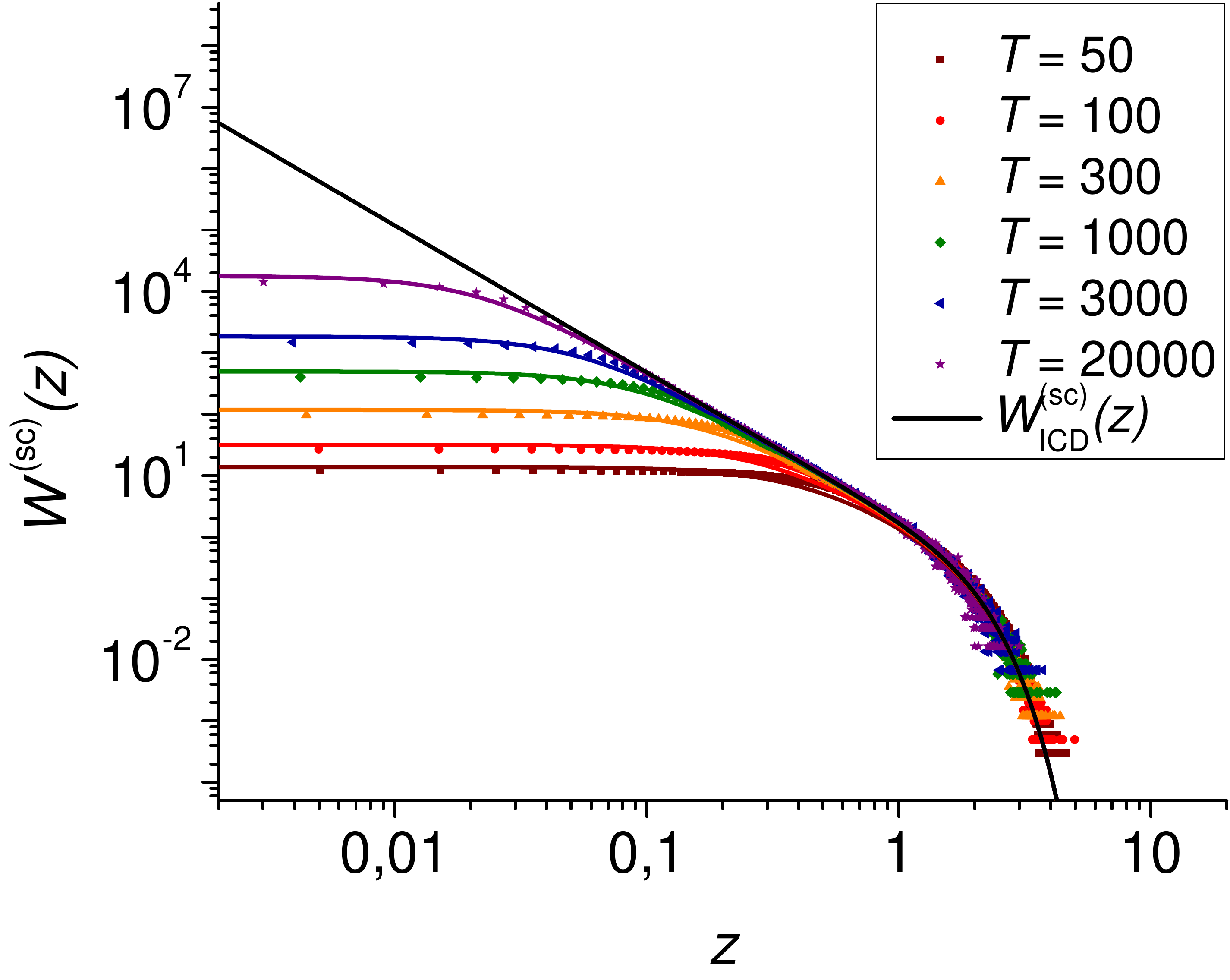}}
\put(308,241){(c)}
\put(155,0){\includegraphics[width=0.28\textwidth]{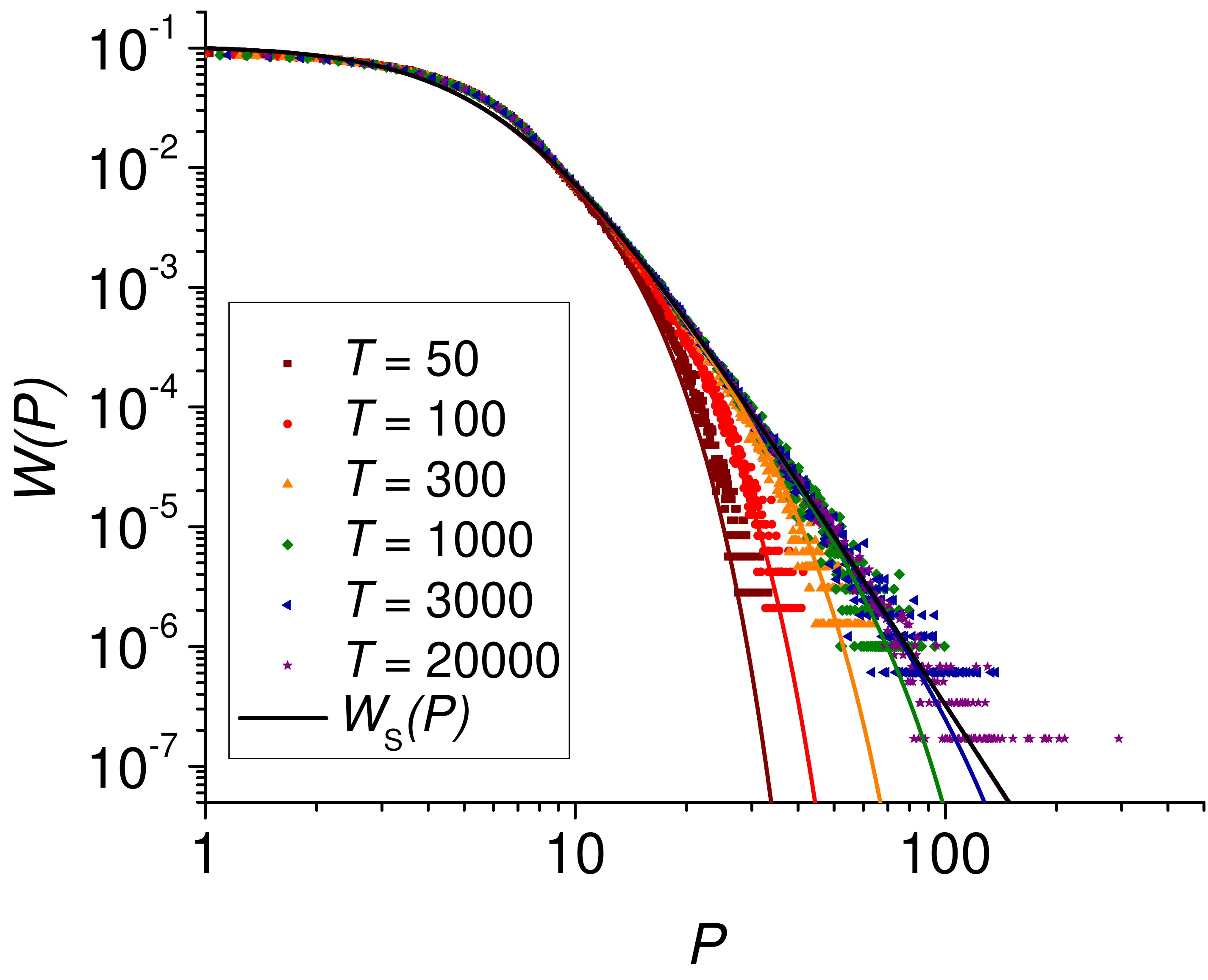}}\put(300,0){\includegraphics[width=0.28\textwidth]{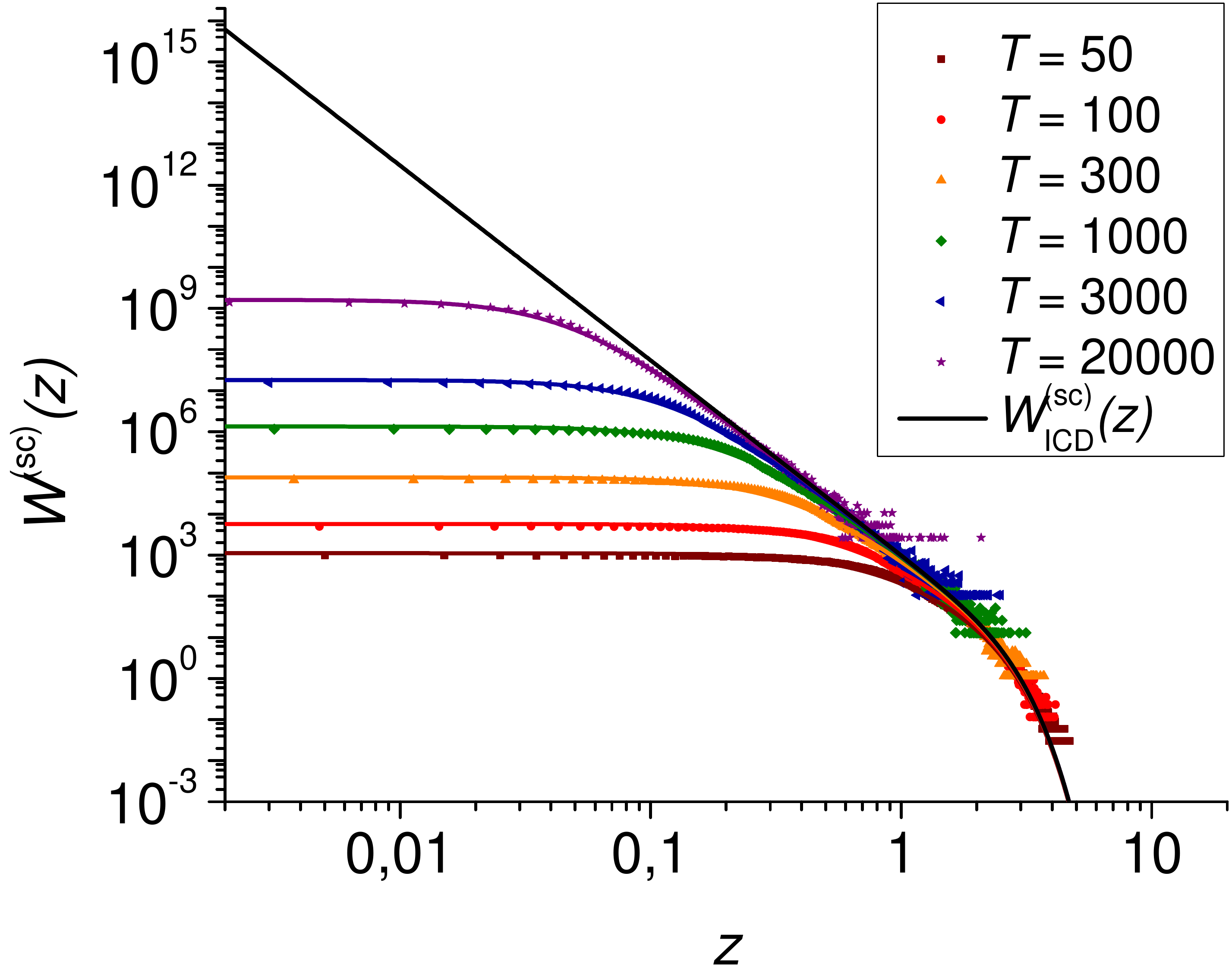}}	
	\end{picture}
	\caption{(a) Phase diagram of the optical lattice. For deep potentials $(q< 2\alpha -2)$ the $q$th moments are determined by the Boltzmann-Gibbs density \eqref{5}, while for shallow lattices $(q>2\alpha -2)$ they are determined by the infinite covariant density \eqref{6}. The autocorrelation function $C_P(T,T_0)$, Eq.~\eqref{10}, is (non)stationary, when the scaling exponent $\alpha>2 \, (\alpha<2)$. (b) Momentum distribution  of an ensemble of $5\cdot10^6$ atoms as a function of time $T$ for $U_0=40\,\Er$ (top), $U_0=80\,\Er$ (bottom) and   $\delta=-10\,\Gamma$. Dots show the results of the Monte Carlo simulations and lines the analytical expression \eqref{7a}. The asymptotic Boltzmann-Gibbs density  \eqref{5} is plotted in black. (c) Scaling form $W^{\rm (sc)}(z)$, Eq.~\eqref{7c}, as a function of the scaling variable $z=P/\sqrt{T}$ for the same parameters and data. The infinite covariant density  \eqref{6} with its characteristic singularity at the origin is plotted in black.} 
	\label{fig:ICD}
\end{figure*}

To analyze the asymptotic solutions of the Fokker-Planck equation \eqref{3}, it is convenient to introduce the dimensionless momentum $P=p/\pr$ and time $T=t\Gamma'$. The infinite time solution, corresponding to the Boltzmann-Gibbs probability   density, is given by \cite{Lut03},
\begin{align}
W_{\rm S}(P) \!=\! \frac{1}{Z} \!\left( 1+ \frac{P^2}{\phi^2} \right)^{\frac{1}{2}-\alpha}\!\!\!\!\!\!\!\!, \ \  Z \!=\!  \frac{\sqrt{\pi}\phi\,\Gamma(\alpha\!-\!1)}{\Gamma\left(\alpha\!-\!1/2\right)},  \; \alpha > 1\label{5}.
\end{align}
On the other hand, the large but finite time solution is an infinite density -- called the infinite covariant density (ICD) in Ref.~\cite{Kes10} to stress its scaling form --  and reads,
\begin{align}
 W_{\text{ICD}}(P,T) 
 \!\simeq\! \left\lbrace
\begin{array}{l@{\,}l}
\!\!\!\frac{1}{Z \Gamma(\alpha)} \left(\frac{P}{\phi}\right)^{1-2\alpha} \!\!\Gamma \left( \alpha,\frac{\chi P^2}{T} \right) &,  \alpha > 1 \\[2ex]
\!\!\!\frac{1}{\Gamma(1-\alpha)} \left(\frac{T}{\chi}\right)^{\alpha - 1} \!\!P^{1-2\alpha} e^{-\frac{\chi P^2}{T}} &, \alpha < 1 .
\end{array} \right. \label{6}
\end{align}
Here $\Gamma(a)$ and $\Gamma(a,z)$ denote the Gamma and incomplete Gamma functions. The three dimensionless parameters $\alpha$, $\chi$ and $\phi$ are given by,
\begin{align}
\alpha &=\frac{\gamma p_c^2}{2D_1}+\frac{1}{2}=\frac{5 U_0}{164\Er}+\frac{1}{2}\,,\ \, \chi =\frac{\Gamma'p_r^2}{4D_1}=\frac{45}{82}\, ,\nonumber \\
\phi &=\frac{p_c}{p_r}\sqrt{1+\frac{D_2}{D_1}}=-\frac{\sqrt{1+\frac{90}{41}\left(\frac{\delta}{\Gamma}\right)^2}}{12 \delta/\Gamma}\frac{U_0}{\Er}\,.
\label{7}
\end{align}
The ICD \eqref{6} diverges like $P^{1-2\alpha}$ at the origin and is hence not normalizable. However, it will prove useful to determine the asymptotic properties of the system.
The exponent $\alpha$, which measures the lattice depth $U_0$ in units of the recoil energy $\Er = \hbar^2 k^2/(2 m)$, is the crucial parameter that  controls the dynamics.  The moments $\langle |P|^q(t) \rangle$ of the Boltzmann-Gibbs  density \eqref{5} diverge for $q > 2 \alpha - 2$; they can thus not be employed  to describe the asymptotic behavior. This divergence was experimentally observed in Ref.~\cite{kat97}.  By contrast, the moments evaluated  via the ICD \eqref{6} are finite for $q > 2 \alpha - 2$ and have been shown to correctly describe the asymptotic time dependence of the system \cite{Kes10,dec11a}. Interestingly, its moments diverge for  $q < 2 \alpha - 2$, when those of the Boltzmann-Gibbs density are finite. The two distributions are hence  complementary: the Boltzmann-Gibbs density describes the properties of the atoms for deep lattices and the ICD those for shallow lattices, see Fig.~\ref{fig:ICD}(a).

Equations \eqref{5} and \eqref{6} can be combined to obtain an approximate time-dependent solution that interpolates between the two, and hence correctly determines the asymptotic behavior of all  moments. It is given by \cite{Kes10},
\begin{align}
W_{\rm app}(P,T) &= \frac{1}{Z} \left( 1+ \frac{P^2}{\phi^2} \right)^{\frac{1}{2}-\alpha} \frac{\Gamma \left( \alpha,\frac{\chi P^2}{T} \right)}{\Gamma(\alpha)} . \label{7a}
\end{align}
This approximate solution is regular at $P=0$, contrary to the ICD, and will thus be used to fit the numerical data below. In the limit $T\to\infty$ it coincides with the stationary solution \eqref{5}. The signature of the ICD \eqref{6} can be revealed  by  introducing the scaling form $W^{\rm (sc)}(z):=T^{-\frac{1}{2}+\alpha}W(z,T)$ of the momentum distributions with scaling variable $z = P/\sqrt{T}$. We then have,
\begin{align}
W^{\rm (sc)}_{\rm app}&(z)= \frac{1}{Z} \left(\frac{1}{T}+\frac{z^2}{\phi^2}\right)^{\frac{1}{2}-\alpha} \frac{\Gamma(\alpha,\chi z^2)}{\Gamma(\alpha)} \label{7b}\\
						&	\xrightarrow[T\to\infty]{} \frac{\phi^{2\alpha-1}}{Z} z^{1-2\alpha} \frac{\Gamma(\alpha,\chi z^2)}{\Gamma(\alpha)}=W^{\rm (sc)}_{\rm ICD}(z) \label{7c}.
\end{align}
In the limit of long times, the scaling form of the approximate solution $W^{\rm (sc)}_{\rm app}(z)$ becomes independent of $T$ and reduces to the scaling form $W^{\rm (sc)}_{\rm ICD}(z)$ of the ICD,  with its characteristic non-normalizable divergence at the origin.

The ICD can also be used to compute the asymptotic two-time momentum correlation function for $T > T_0$ \cite{dec13},
\begin{align}
C_P(T,T_0) \!\simeq\! \frac{\sqrt{\pi}}{\Gamma(\alpha\!+\!1)} \!\left\lbrace 
\begin{array}{@{}l@{}l}
\frac{\phi^{2 \alpha - 1}}{Z \Gamma(\alpha)} \left(\frac{T_0}{\chi}\right)^{2-\alpha} \!\!\!f_{\alpha} \!\left( \frac{T-T_0}{T_0} \right) &, \alpha > 1 \\[2ex]
\frac{\sqrt{\pi}}{\Gamma(1 - \alpha)} \frac{T_0}{\chi} \, g_{\alpha} \!\left( \frac{T-T_0}{T_0} \right) &, \alpha < 1.
\end{array} \right.  \label{8}
\end{align}
The two functions $f_{\alpha}(s)$ and $g_{\alpha}(s)$ are here defined as,
\begin{align}
f_{\alpha}(s) &\!=\! s^{2-\alpha} \!\!\int_{0}^{\infty} \!\!{\rm d} y \ y^{2} e^{-y^2} {}_{1}\text{F}_{1} \!\left( \frac{3}{2}, \alpha + 1, y^2 \right) \Gamma \!\left( \alpha, y^2 s \right), \nonumber \\
g_{\alpha}(s) & \!=\!  s^{2-\alpha} \!\!\int_{0}^{\infty} \!\!{\rm d} y \ y^{2} e^{-y^2} {}_{1}\text{F}_{1} \!\left( \frac{3}{2}, \alpha + 1, y^2 \right) e^{-y^2 s}
\end{align}
where ${}_{1}\text{F}_{1}(x)$ is a hypergeometric function. The correlation function \eqref{8} depends in general on both $T$ and $T_0$, and is thus nonstationary and exhibits aging. However, it  has a stationary limit, $T \rightarrow \infty$, for $\alpha > 2$ ($U_0 > 49.2\,  \Er$) which depends only on the time lag $T - T_0$,
\begin{align}
C_{P,\text{s}} (&T-T_0) \simeq \phi^{2 \alpha - 1} \frac{\pi \, \Gamma(\alpha - 2) }{4 Z \, \Gamma^2\!\left( \alpha - \frac{1}{2} \right) } \left(\frac{T - T_0}{\chi}\right)^{2-\alpha}\!\!\! \label{10}.
\end{align}
Expression \eqref{10} was  obtained  using the stationary solution of the Fokker-Planck equation in Ref.~\cite{Mar96}. For $\alpha < 2$, no stationary limit exists as the second moment of the Boltzmann-Gibbs density \eqref{5} is infinite. In this regime, Eq.~\eqref{8} can only be calculated via the ICD \eqref{6}.

\textit{Monte Carlo simulations.}
We use a Runge-Kutta algorithm of second order to integrate the Langevin equations corresponding to the phase space equations \eqref{1}. The  discrete form of these Langevin equations reads \cite{Pet99}
\begin{align}
x(t+&\dt)=  x(t)+\frac{p(t)}{m}\dt,\label{12}\\
p(t+&\dt)=   p(t) + \left\{ \begin{array}{@{}l@{}l}
           &-U_{\mp}'(x)\dt +\eta\sqrt{2D_{\pm\mp}(x)/\gamma_{\pm\mp}(x)}  \\
           &\qquad \text{if }\rho<\gamma_{\pm\mp}(x)\dt\\[2 ex]
           &-U_{\pm}'(x)\dt + \eta\sqrt{2D_{\pm\pm}(x)\,\dt} \\
           &\qquad \text{if }\rho\geq\gamma_{\pm\mp}(x)\dt,
       \end{array}\right.\nonumber
\end{align}
where $\eta$ is a random number drawn from a standard normal distribution and $\rho$ is a random number equally distributed between zero and one. Equations \eqref{12} have been successfully used to investigate the evolution of cold atoms in deep lattices \cite{Pet99}. However, in shallow lattices they lead to infinitely large momentum jumps at the nodes of the potential where $\gamma_{\pm\mp}(x)\sim 1 \pm \cos(2 k x)=0$ and $D_{\pm\mp}(x)\sim 6 \mp \cos(2 k x) \neq 0$ (see Eqs.~(14)). These divergences are unphysical and follow from the semiclassical approximation: in the quantum description, atoms are never exactly localized at the nodes of the potential. To remove the divergences, we modify  the parametrization such that  $D_{\pm\mp}(x) \propto\gamma_{\pm\mp}(x)$ \cite{Cas94}. More precisely, by choosing $D_{\pm\mp}(x)=  (6\hbar^2 k^2 \Gamma'/90) [1 \pm \cos(2 k x)]$,  the averaged diffusion coefficients in Eq.~\eqref{3}, hence the Fokker-Planck dynamics, remain unchanged. A similar approach has been shown to yield good agreement between 2D semiclassical and quantum Monte Carlo simulations \cite{Cas94}.

In our simulations, the atoms were initially equally distributed over one lattice period $\pi/k$ and each trajectory started with zero momentum, as assumed in the derivation of the ICD \cite{Kes10}. From an experimental point of view, it suffices that the initial momentum distribution is much narrower than the stationary density, that is, the atoms need to be cold enough. We extracted the  three dimensionless parameters \eqref{7} of the system  from the numerics for an ensemble of $10^7$ atoms and a simulation time of $T=t\Gamma'=20000$. The determination of the exponent $\alpha$ of the steady state distribution \eqref{5}  is quite involved since the power-law behavior is limited to a finite momentum interval. We used the  very accurate maximum likelihood estimation (MLE) technique  described in Ref.~\cite{Bar12}. The two remaining parameters $\phi$ and $\chi$ were  obtained using a standard least-squares fit  of the distribution $W_\text{app}(P,T=20000)$ to the data. Details of  the fitting procedure, as well as typical fit results, can be found in the Supplemental Material.
For the scaling exponent $\alpha$, we find excellent agreement (less than $3 \%$ deviation)  between the fitted values and the analytical expressions \eqref{7} obtained from the approximate Fokker-Planck equation \eqref{3} (see Fig.~4 in the Supplemental Material). The parameters $\phi$ and $\chi$  display the predicted linear and constant dependences on $U_0$, albeit with somewhat larger deviations (between $30\%$ and $37\%$ for $\phi$ and below $13\%$ for $\chi$). 
These findings indicate that the diverse approximations entering the Fokker-Planck equation, in particular the crude spatial averaging, preserve the important scaling properties of the microscopic dynamics \eqref{1} and only affect  the value of  numerical prefactors.

 Figure \ref{fig:ICD}(b) shows a comparison of the simulated momentum distribution (for $5\cdot10^6$ atoms)  with the analytical approximate solution \eqref{7a}, using the fitted parameters $\alpha$, $\phi$ and $\chi$, for different evolution times $T\in[50,20000]$ and different lattice depths, $U_0=40\,\Er$ (top) and $U_0=80\,\Er$ (bottom). We observe  excellent agreement in the range of validity of the ICD (large $P$ and large $T$) as expected, but also for moderate values of $P$ and $T$; deviations are only seen for small momenta, $P\lesssim10$, and small  times, $T\lesssim 100$.
  As the evolution time increases, the data  converge towards the Boltzmann-Gibbs  density \eqref{5}. The scaling form, $W^{\rm (sc)}(z)=T^{-\frac{1}{2}+\alpha}W(z,T)$,  is plotted in Fig.~\ref{fig:ICD}(c).  For large $z$ and  long evolution times $T$, the rescaled data loose their explicit time dependence, as predicted, and clearly converge to the scaling form $W^{\rm (sc)}_{\rm ICD}(z)$ of the ICD given by Eq.~\eqref{7c}. For small $z$, the data are well described by the approximate solution \eqref{7b}.
We found the above results to be independent of the specific values of  potential depth and detuning within the investigated parameter space: $U_0/\Er\in[40,80]$ for constant $\delta=-10\,\Gamma$, and $\delta/\Gamma\in[-5,-30]$ for constant $U_0=50\,\Er$.

\begin{figure}[htbp]
\centering
\begin{picture}(150,250)
\put(-25,250){(a)}	\put(-15,130){\includegraphics[width=0.33\textwidth,trim=12mm 4mm 40mm 10mm,clip]{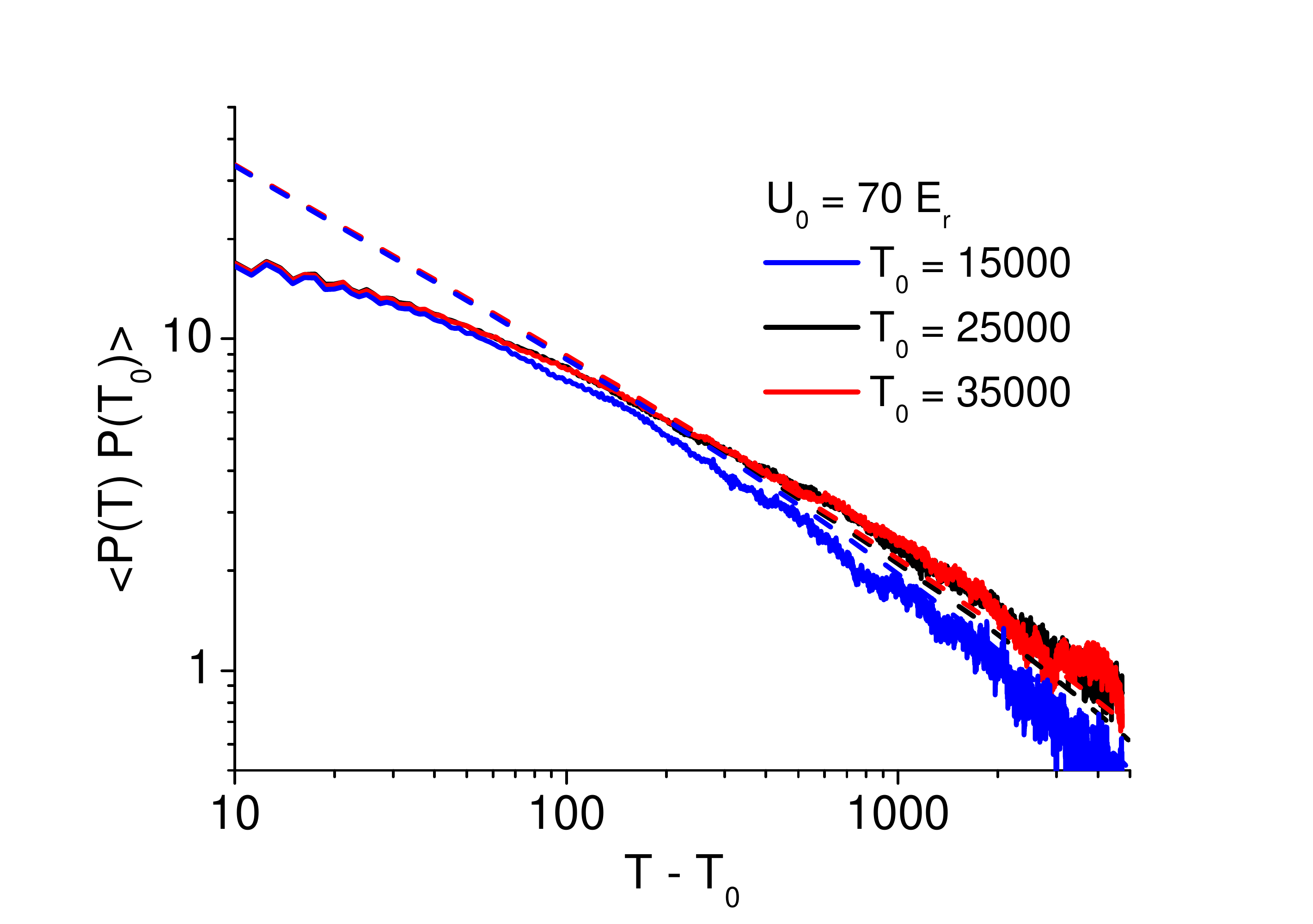}	\hspace{-.5cm}}
\put(-25,120){(b)}	\put(-15,0){\includegraphics[width=0.33\textwidth,trim=12mm 4mm 40mm 10mm,clip]{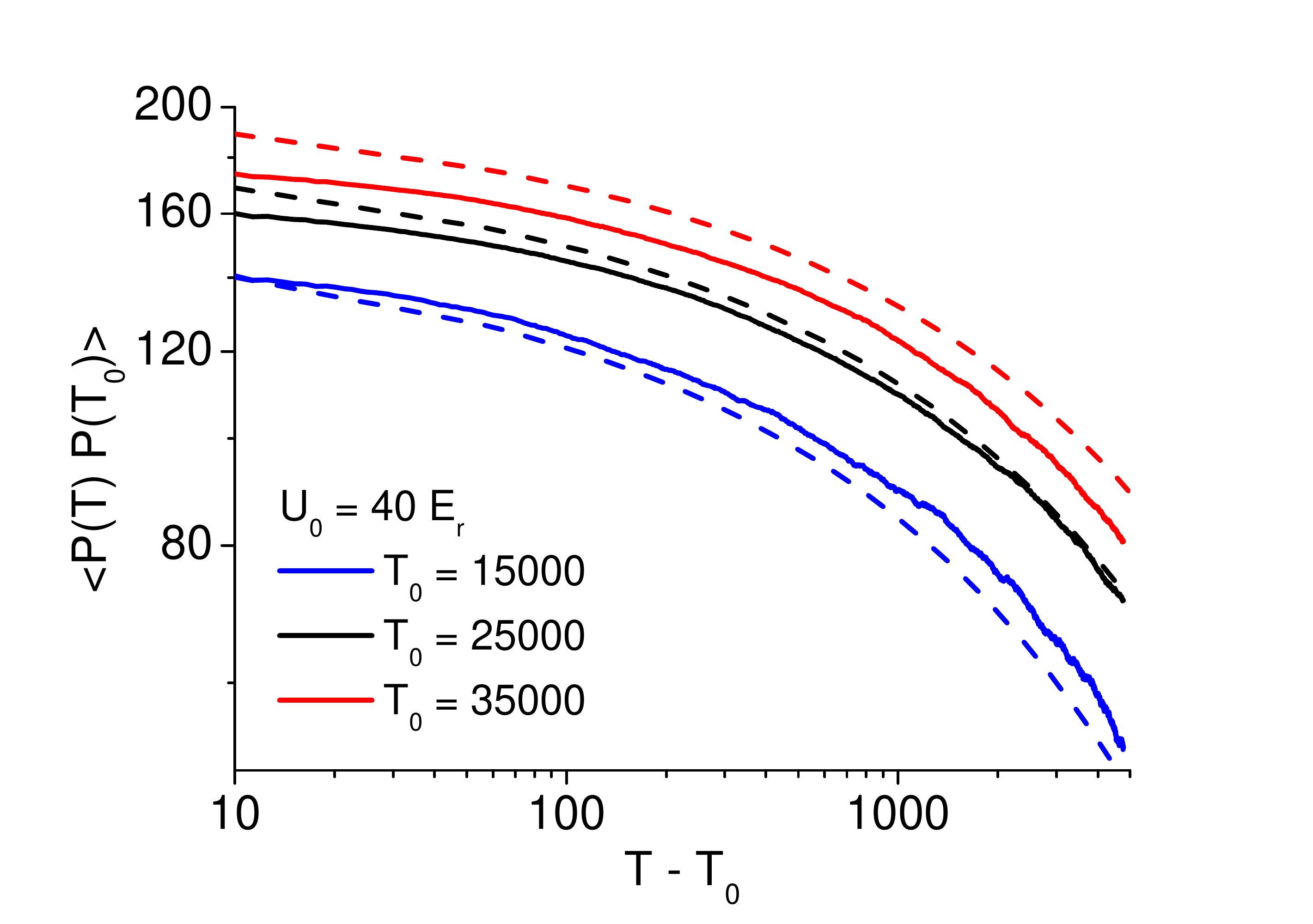}}
	\end{picture}
	\caption{Momentum correlation function $C_P(T,T_0)$ in the (a) stationary (deep lattice, $U_0=70 E_R$) and (b) nonstationary (shallow lattice, $U_0=40 E_R$) regime for $2 \cdot 10^5$ trajectories. Blue, black and red lines correspond, respectively, to $T=15000$, $T=25000$ and $T=35000$. Solid lines are the results of the semiclassical Monte Carlo simulations, while  dashed lines show the analytical expression \eqref{8} with  fitted parameters.} 
	\label{fig:corr}
\end{figure}

The two-point momentum correlation  function $C_P(T,T_0)=\langle P(T)P(T_0)\rangle$ is plotted in Fig.~\ref{fig:corr}. It is quasistationary for deep lattices, Fig.~\ref{fig:corr}(a), exhibiting little dependence on the initial time $T_0$, provided the latter is large enough. The oscillations seen at short time differences $T-T_0$ stem from the periodicity of the optical lattice; they disappear for longer time lags. For shallow lattices, Fig.~\ref{fig:corr}(b), the nonstationarity is clearly visible: the magnitude of the correlation function increases with $T_0$, as predicted by Eq.~\eqref{10}. While good agreement between numerics and analytics, Eq.~\eqref{10}, is found in the stationary regime, it is only qualitative in the nonstationary regime. The latter can be traced back to the assumption of stationarity of  the population difference, $\varphi(x,p,t) = W_{+}(x,p,t) - W_{-}(x,p,t)$, in the derivation of the Fokker-Planck equation \eqref{3}. A more accurate  description of the nonstationary properties of the system would thus require the inclusion of the relaxation of the function $\varphi(x,p,t)$.

\textit{Conclusions.}
We have performed semiclassical Monte-Carlo simulations of the microscopic dynamics of cold atoms in shallow dissipative optical lattices and compared it to the prediction of an approximate Fokker-Planck equation in the asymptotic, finite time regime. We have  found that the scaling behavior  is the same in both approaches and that only numerical prefactors slightly differ. We have further shown that the infinite density, which determines the properties of the system in the regime where the Boltzmann-Gibbs distribution fails, is observable even for moderate evolution times. In addition, we have demonstrated that the Fokker-Planck equation provides a good description of the autocorrelation function in the stationary phase and an approximate description in the nonstationary phase.  An experimental confirmation of the presence of the infinite density in optical lattices -- with its non-normalizable asymptotic divergence at the origin -- would constitute a significant step forward in the investigation of nonergodic systems. 

We thank Eli Barkai for discussions and acknowledge support from the DFG (contract No LU1382/4-1).

\bibliography{ICD_BIB}

\section{Supplemental Material}

\subsection{Microscopic coefficients}
The position-dependent microscopic coefficients in the phase space equations \eqref{1} are given by \cite{Pet99}, 
\begin{align}
&U_{\pm}(x) \!=\! \frac{U_0}{2} [-2 \pm \cos(2 k x)], \,  \gamma_{\pm \mp}(x) 
\!=\! \frac{\Gamma'}{9} [1 \pm \cos(2 k x)] \nonumber, \\
& D_{\pm \pm}(x) \!=\! \frac{7 \hbar^2 k^2 \Gamma'}{90} [5 \pm \cos(2 k x)] \nonumber, \\
& D_{\pm \mp}(x) \!= \!\frac{ \hbar^2 k^2 \Gamma'}{90} [6 \mp \cos(2 k x)] . \label{2}
\end{align}

\subsection{Details of the fitting procedure}
\begin{figure}[htbp]
	\centering
	\includegraphics[width=0.35\textwidth]{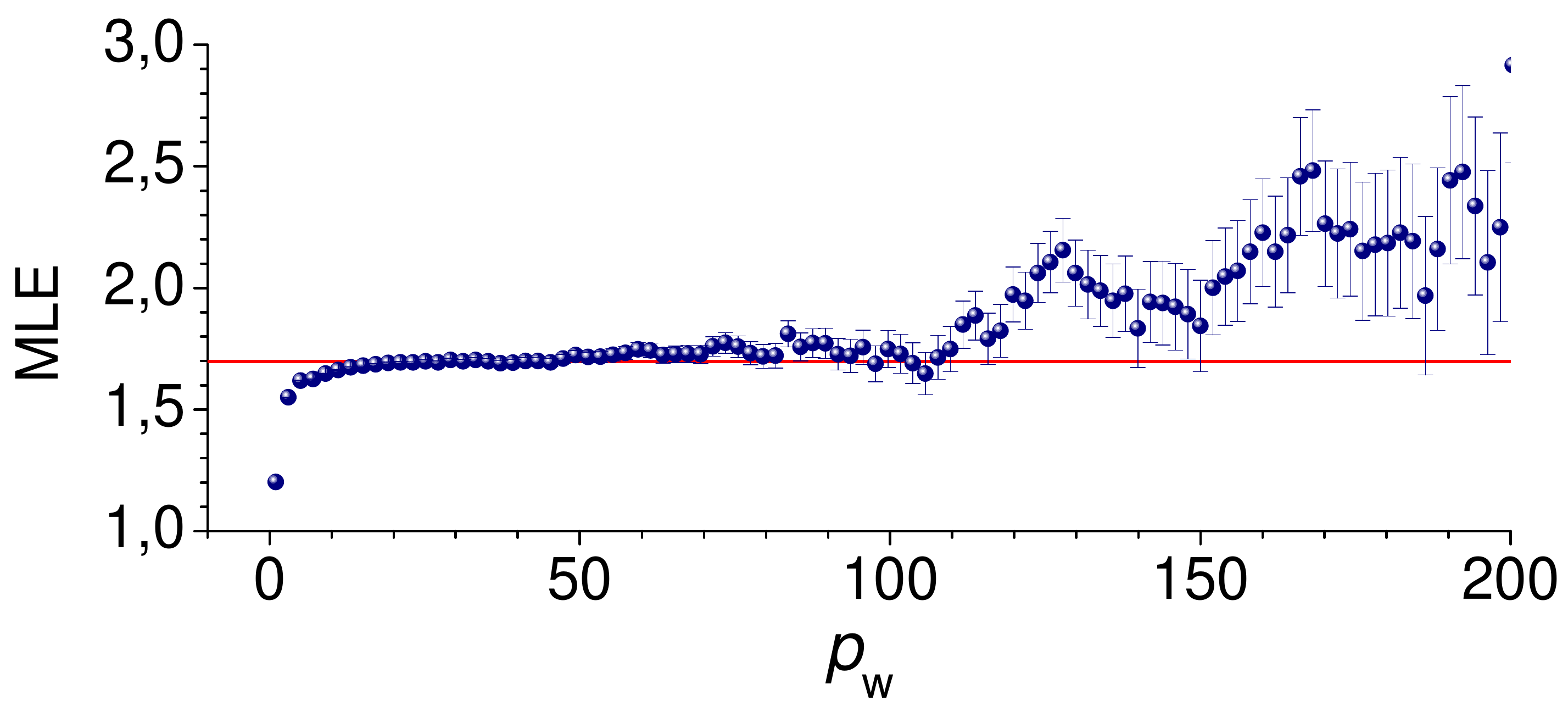}
	\caption{Typical plot generated in the MLE analysis of the exponent $\alpha$. The MLE value calculated from all data points within a momentum window plotted against the window position $P_\text{w}$. The value of $\alpha$ indicated by the red solid line is determined by averaging over the plateau region.} 
	\label{fig:MLE}
\end{figure}

\begin{figure}[htbp]
	\centering
	\includegraphics[width=0.27\textwidth]{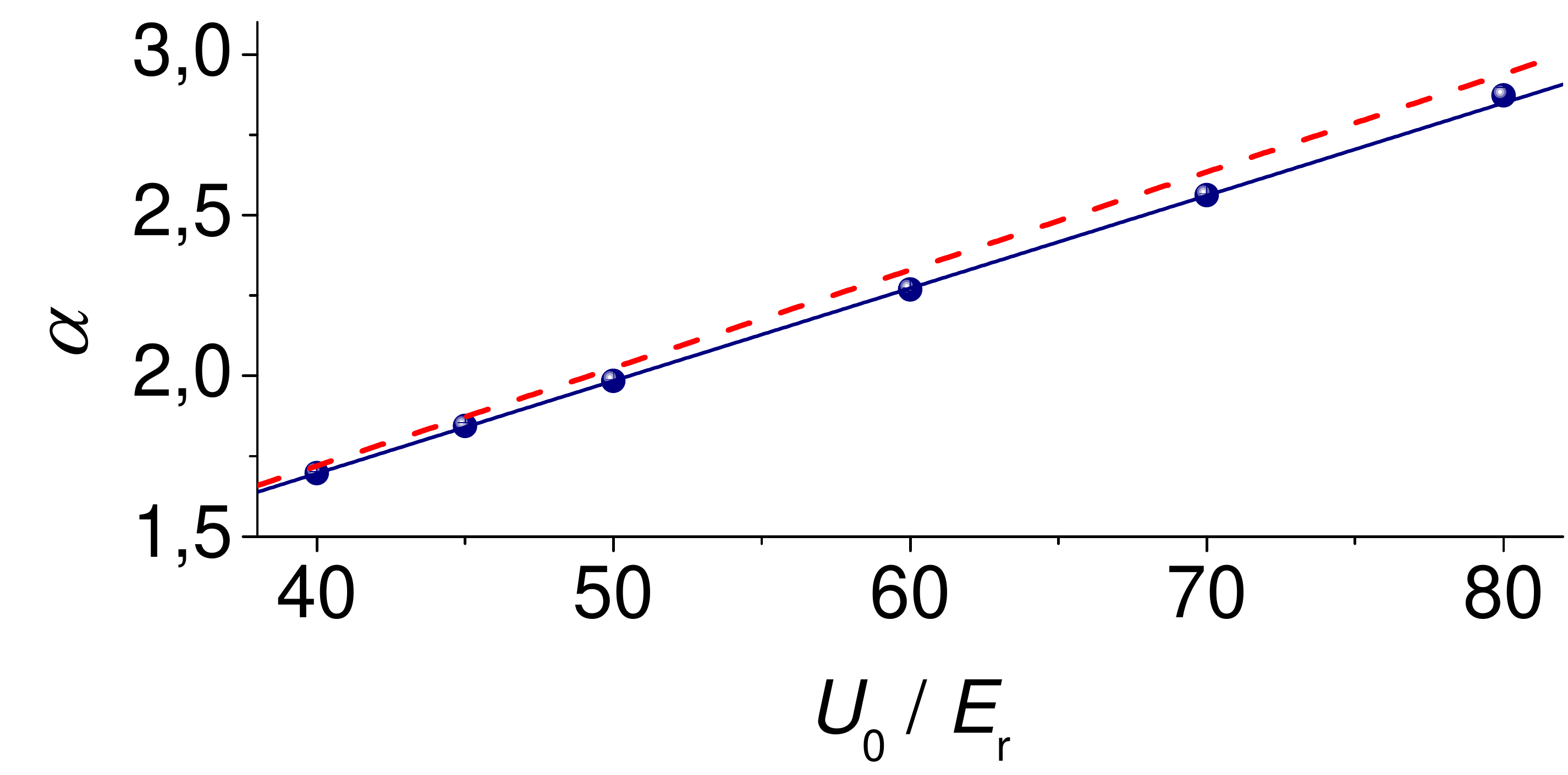}\\
	\includegraphics[width=0.27\textwidth]{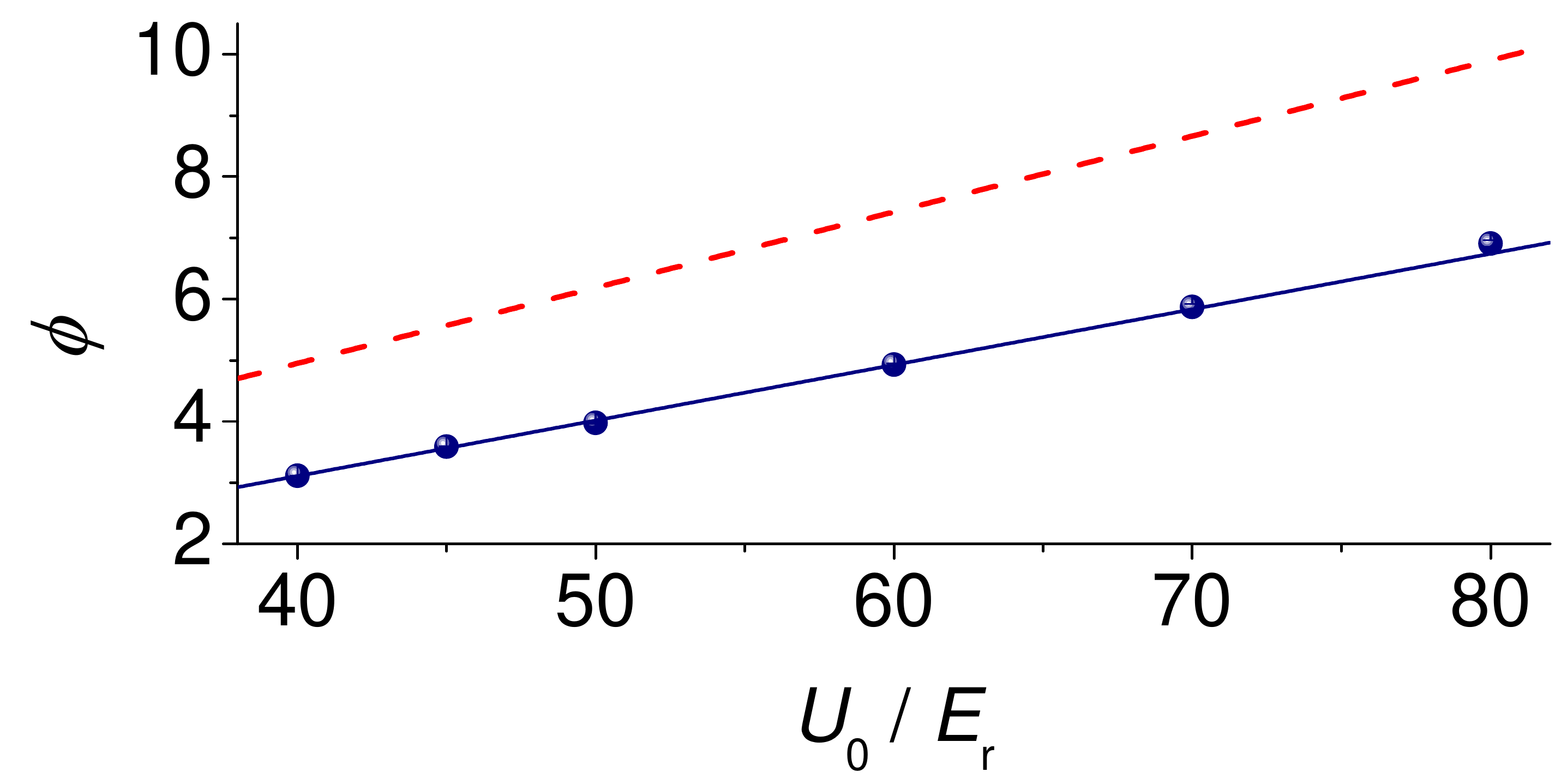}\\ \includegraphics[width=0.27\textwidth]{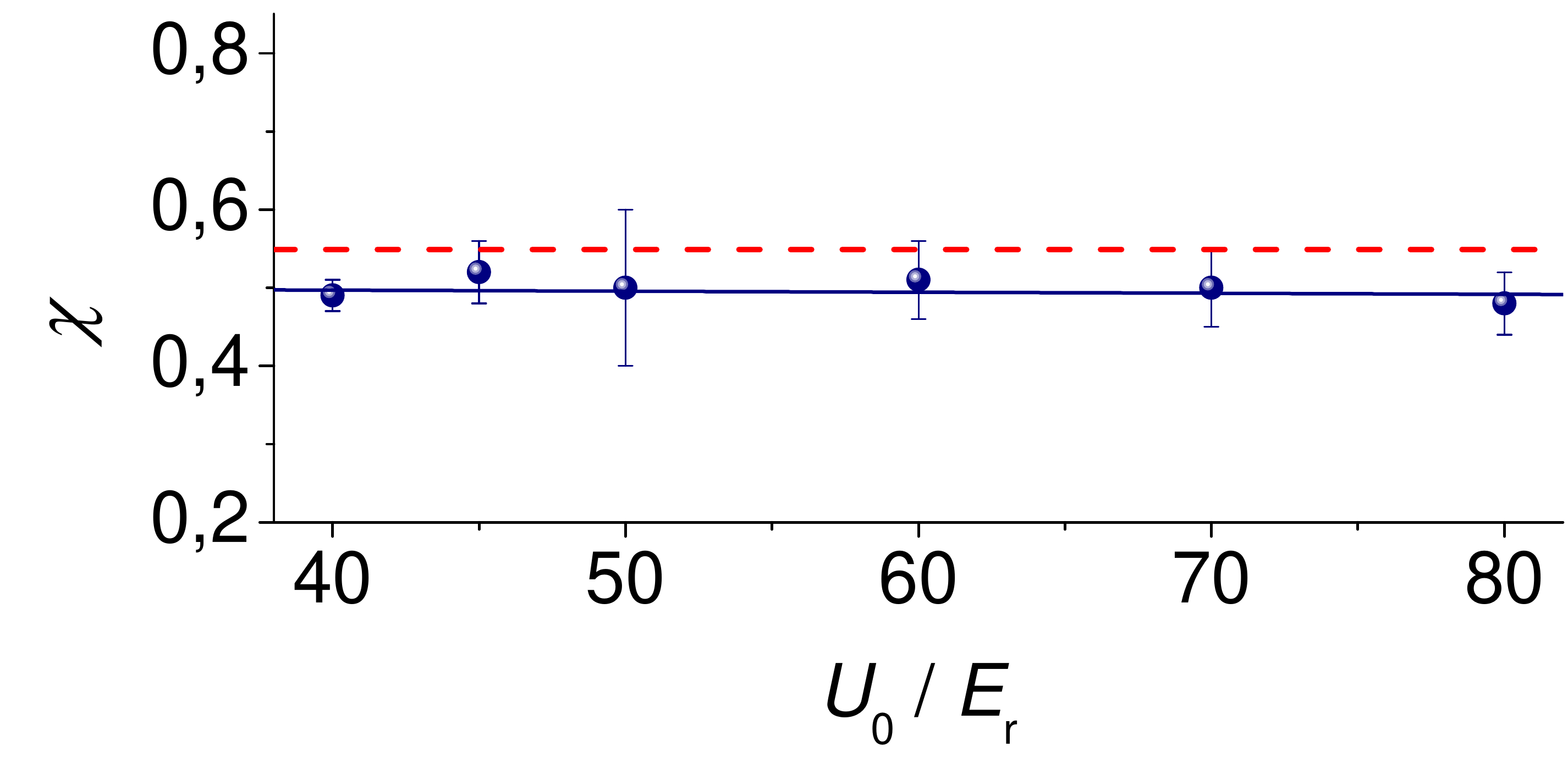}
	\caption{Parameters $\alpha$, $\phi$ and $\chi$ , Eqs.~\eqref{7}, as a function of the  potential depth $U_0$  at a detuning $\delta=-10\,\Gamma$. The dashed red lines show the theoretical dependence given by Eq.~\eqref{7}, the solid blue lines are a fit to the data.} 
	\label{fig:par-summary}
\end{figure}

To evaluate the three parameters $\alpha$, $\varphi$ and $\chi$, Eqs.~\eqref{7}, from the numerics we first determine the exponent $\alpha$ of the steady state distribution \eqref{5} using the maximum likelihood estimation (MLE) method  described in Ref.~\cite{Bar12}. We calculate the MLE value for the parameter $\alpha$ including all final momentum values of the simulated trajectories that lie within a momentum window of several tens of \pr width. By shifting this window along the momentum axis, we scan the momentum distribution and can identify the region in which it follows a power law by plotting the MLE value versus the window position $P_\text{w}$. The existence of a pronounced plateau in this plot indicates that the data follow a power law. We use an average over the plateau region to determine $\alpha$. A typical result of this procedure is shown in Fig.~\ref{fig:MLE}.
The two remaining parameters $\varphi$ and $\chi$ are obtained using a standard least-squares fit  of the distribution $W_\text{app}(P,T=20000)$ to the data. This works reliably because variations in $\varphi$ shift the whole distribution along the $p$-axis, while variations in $\chi$ determine the position of the cut-off. Because the central parts of the momentum distributions show deviations from the approximate solution, we omit data points at $P$-values smaller than some threshold $P_\text{th}$ in the fit. By varying $P_\text{th}$ and searching for a plateau in a way similar to the MLE-analysis shown in Fig.~\ref{fig:MLE}, we are able obtain reliable results for $\varphi$ and $\chi$. The results of the fitting procedure are shown in Fig.~\ref{fig:par-summary}.

\subsection{Variation of the detuning $\delta$}

\begin{figure}[ht]
	\centering
	\includegraphics[width=0.28\textwidth]{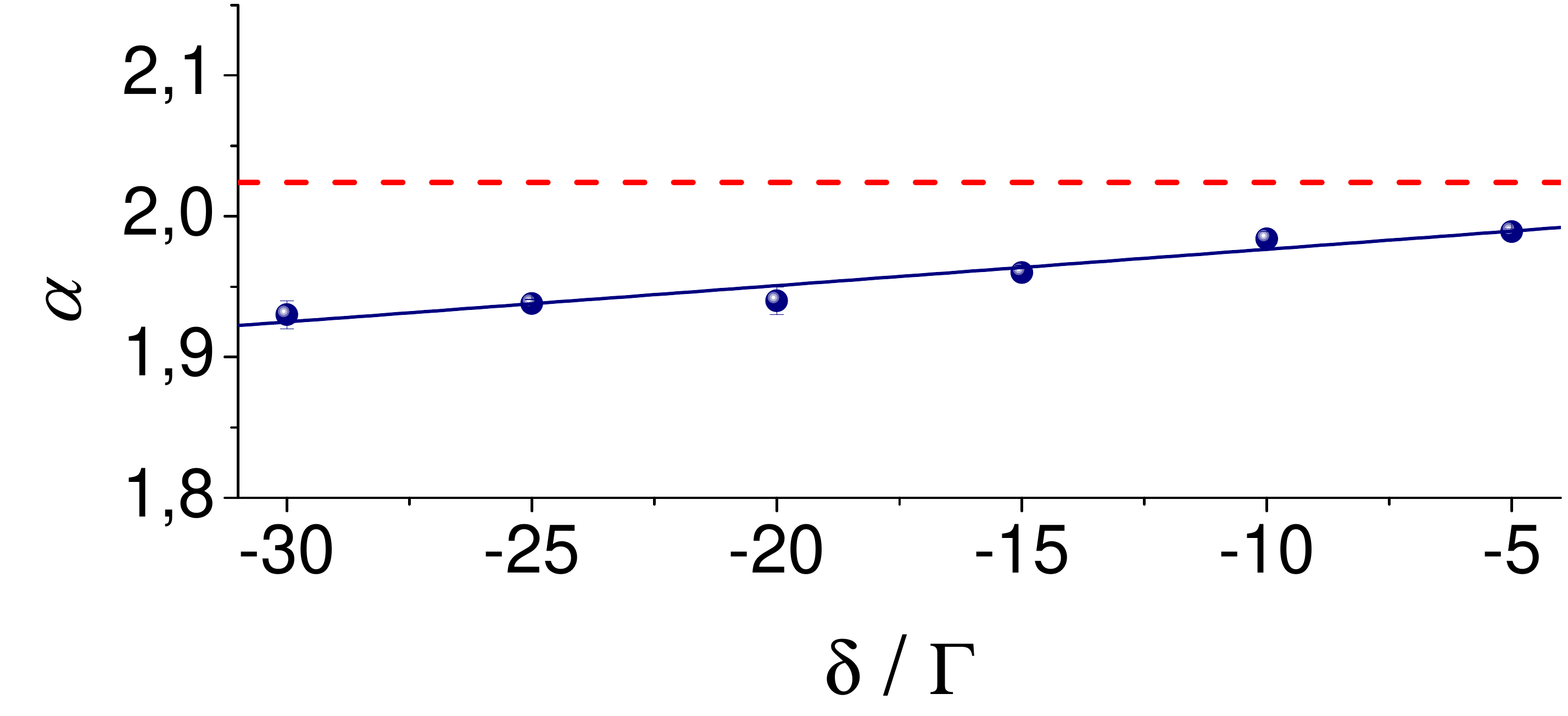}\\
	\includegraphics[width=0.28\textwidth]{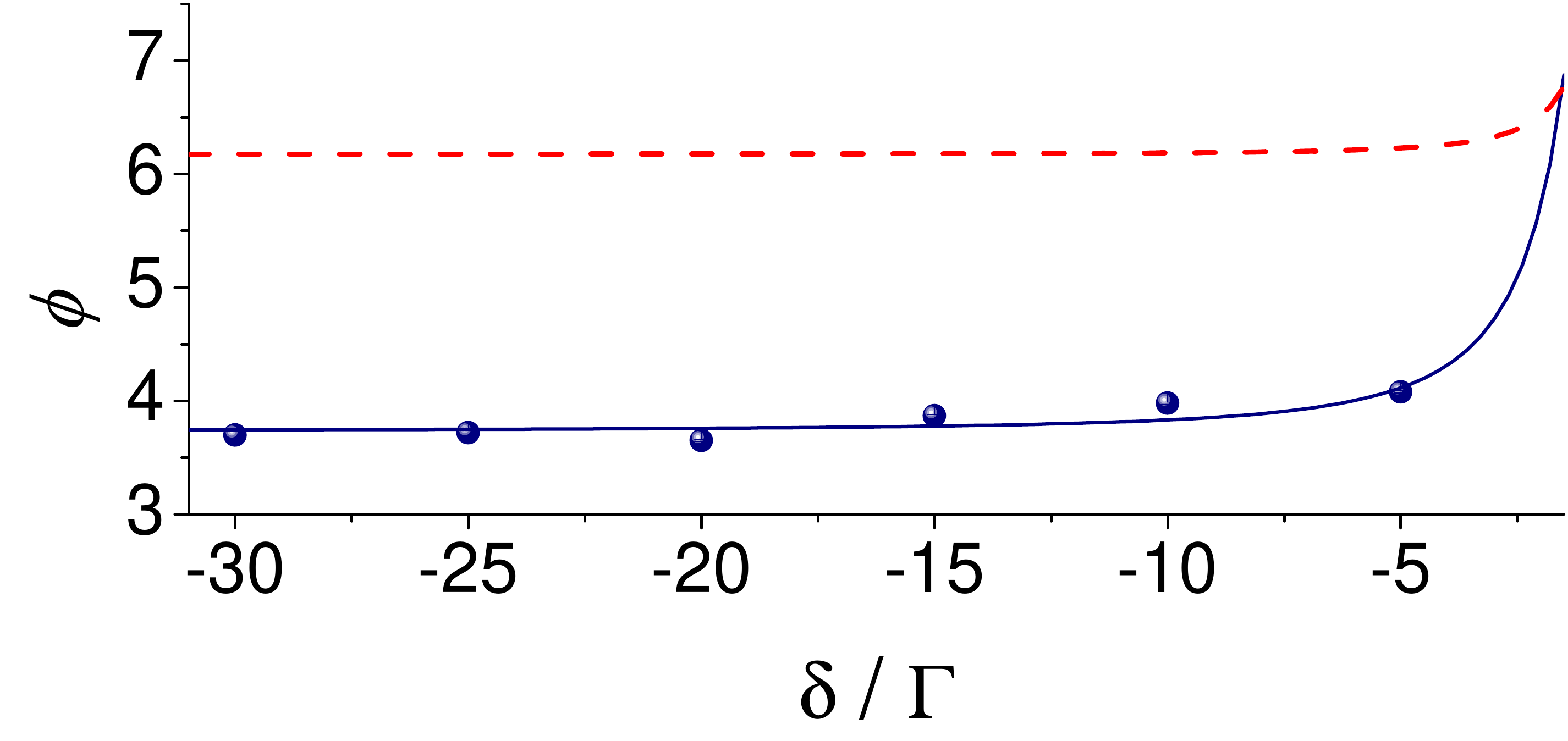}\\\includegraphics[width=0.28\textwidth]{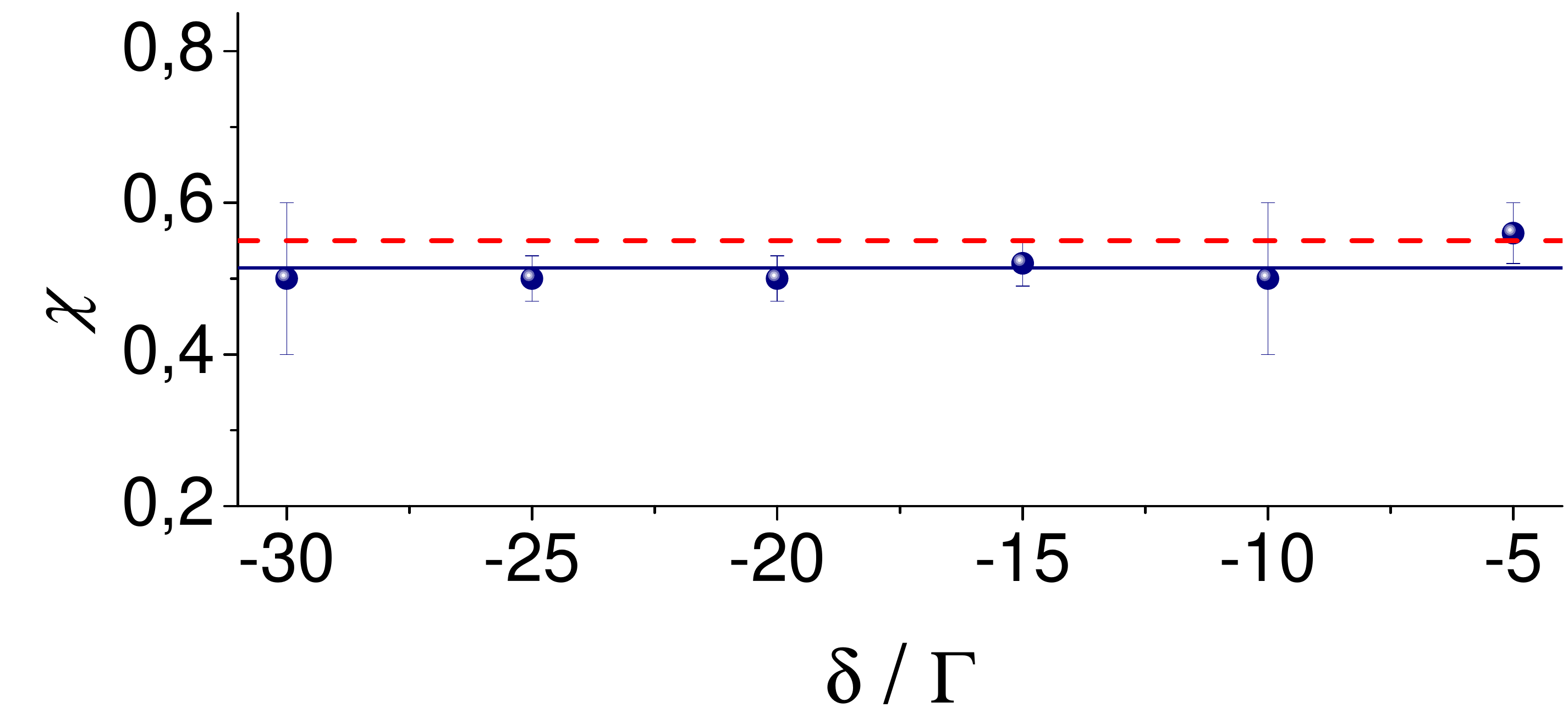}
	\caption{Parameters $\alpha$, $\phi$ and $\chi$, Eqs.~\eqref{7}, as a function of the detuning $\delta$ at a potential depth $U_0=50\,\Er$. The dashed red lines show the theoretical dependence given by Eq.~\eqref{7}, the solid blue lines are a fit to the data.} 
	\label{fig:par-delta-summary}
\end{figure}

We have also investigated the change of the momentum distribution under variation of the detuning $\delta$. To this end we have repeated the analysis described above for different detunings $\delta/\Gamma\in[-5,-30]$ at a fixed potential depth $U_0=50\,\Er$. The results of the fitting procedure for the parameters $\alpha$, $\phi$ and $\chi$ are summarized in Fig.~\ref{fig:par-delta-summary}.
\begin{figure}[hb]
	\centering	
	\includegraphics[width=0.245\textwidth]{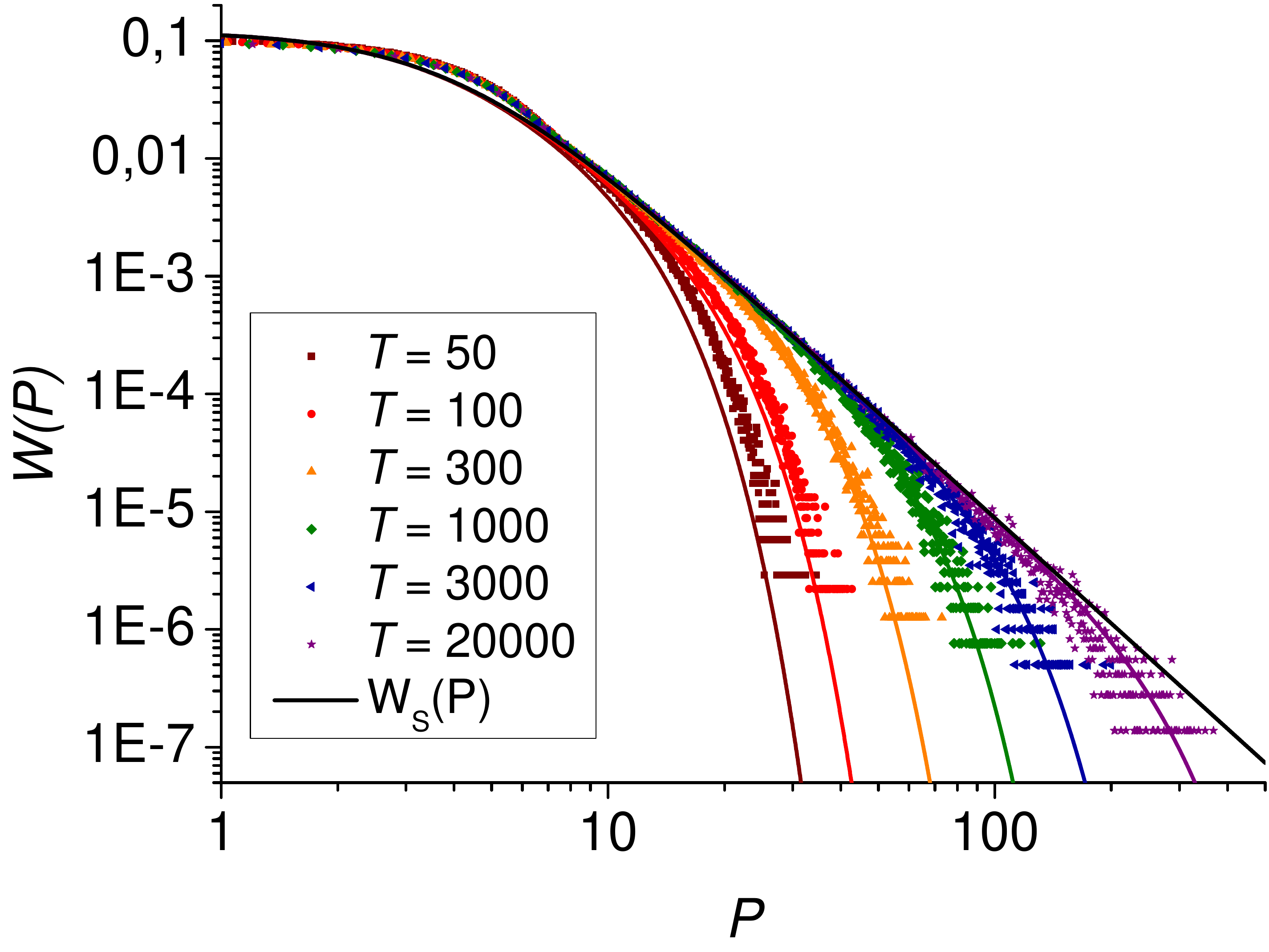}\hspace{-0.2cm}\includegraphics[width=0.245\textwidth]{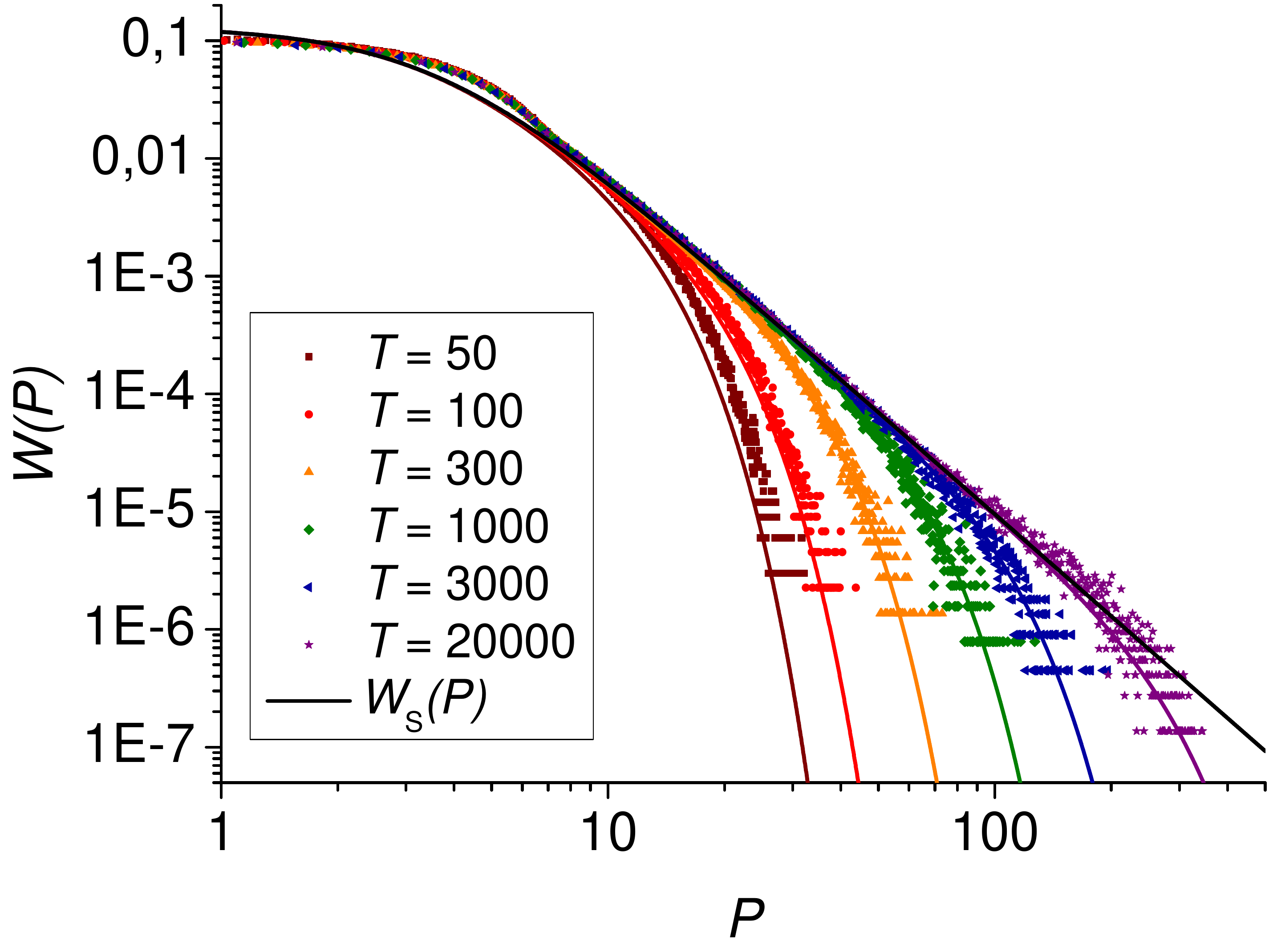}\\	
	\caption{Momentum distribution as a function of  time $T$ for two  detunings, (left) $\delta=-5\,\Gamma$ and (right) $\delta=-20\,\Gamma$, with   a potential depth $U_0=50\,\Er$. The plot for $T=20000$ on the right shows the slightly increased momentum probability density deviating from the expected curves at high momenta. We believe this to be an artefact of the simulations due to an insufficiently resolved lattice structure.} 
	\label{fig:ICD-delta}
\end{figure}
For $\phi$ and $\chi$, we find good qualitative agreement between the fitted values and the theoretical curves given by Eq.~\eqref{7}. Quantitatively, we observe deviations from the theory consistent with our results for varying potential depth shown in Fig.~\ref{fig:par-summary}. We attribute these deviations to the spatial averaging procedure used to derive Eq.~\eqref{3}. In addition, we find a small dependence of the scaling exponent $\alpha$ on the detuning $\delta$. This observation is not supported by the theory which predicts $\alpha$ to be a constant with respect to variations in $\delta$. The deviation can be traced back to the momentum distributions as shown in Fig.~\ref{fig:ICD-delta}. For larger detunings, the data show a slightly enhanced probability $W(P)$ at high momenta $P$ in comparison to the expected analytical approximate solution \eqref{7a}. 
The increased probability leads to an underestimation of the true power-law exponent $\alpha$ by its maximum likelihood estimator. We think that the discrepancy at high $P$ is an artefact produced by the finite size of the time step $\text{d}t$ in the numerics. For the simulations to converge, the time step must fulfil two criteria \cite{Pet99}. First, $\text{d}t\ll 1/\Gamma'$, i.\,e. individual scatter events are resolved. Second, the structure of the optical potential needs to be resolved, i.\,e. $\text{d}t\ll\lambda/(2v_\text{max})$, where $v_\text{max}$ is some assumption on the maximal velocity of particles in the lattice. For the shallow lattices investigated here, the second criterion is more restrictive. In our numerics, we have assumed a maximal velocity of $v_\text{max}=450\,\pr/m$, given by the limit of computer power available. For particles travelling with $p\gtrsim 450\,\pr$, the lattice structure is thus insufficiently resolved and Sisyphus cooling ceases, leading to the observed increase of the momentum probability density. The explicit dependence of the data on the detuning $\delta$ enters the simulations via the scaling of the simulation time in terms of $1/\Gamma'\propto |\delta|$. For increasing values of $|\delta|$, the number of iterations increases and the deviations become more pronounced as seen in Fig.~\ref{fig:par-delta-summary}.

\end{document}